\documentstyle[epsfig]{aipproc}

\begin{document}
\title{On the Nature of Black Hole Entropy}

\author{Ted Jacobson}
\address{Institute for Theoretical Physics, 
University of California, Santa Barbara, CA 93106\\
 Department of Physics, University of Maryland,
College Park, MD 20742} 
%\lefthead{LEFT head}
%\righthead{RIGHT head}
\maketitle

\begin{abstract}
I argue that black hole entropy counts only those
states of a black hole that can influence the outside,
and attempt (with only partial success) to defend
this claim against various objections, all but one coming
from string theory. Implications for the nature 
of the Bekenstein bound are discussed, and in particular
the case for a holographic principle is challenged.
Finally, a generalization
of black hole thermodynamics to ``partial event horizons"
in general spacetimes without black holes is proposed. 
\end{abstract}
 
\section*{Black hole entropy and internal states} 

Let me begin by giving several reasons why we should not think
that the Bekenstein-Hawking entropy 
$S_{\rm BH}=A/4\hbar G$ of a
black hole counts the number of  internal
states of the black hole. (By ``the entropy" of a black hole
I will always mean $A/4\hbar G$ in this article.)
These reasons have been enunciated
in a thoughtful article by  Rafael Sorkin\cite{sorksm},
which I will borrow from here.  
\begin{enumerate} 
\item\label{firstreason}
The spatial region inside a black hole horizon can have
arbitrarily large volume, with room for an arbitrarily
large number of states. For example a  Friedmann universe
of any size can be joined to the interior of a
Schwarzschild black hole. Thus the number of possible
internal states of a black hole is unbounded. 
\item A
black hole is not in ``internal equilibrium",  so why
should its thermodynamic entropy refer to its interior
states? 
\item Conditions inside the horizon are causally
disconnected from the outside, so how can the states inside
be thermodynamically relevant to the outside?  
\item  According to local
quantum field theory the evaporation of a black hole is
unitary, at least until the final stages, and  the Hawking
radiation is  correlated to field degrees of freedom
inside the black hole. The number of internal states of
the black hole must therefore remain large enough to 
store all the correlations maintaining the purity of the
total state. As a black hole evaporates, however, its
area  and therefore its entropy decreases. Thus the
entropy must not be counting the number of internal
states. 
\end{enumerate} 

Regarding point 1, it should be mentioned that the example
given will have a white hole horizon and singularity
in its past (assuming the weak energy condition holds) so
it is not a configuration that would evolve from an
ordinary collapse process\cite{sorwaljiu}. 
It is nevertheless a possible
state of the black hole.  

There is by now a ``standard" argument
against points 3 and 4, namely, that local quantum field
theory may be inapplicable. This argument is suggested by
(but not restricted to) string theory, in which local
quantum field theory is only an approximation valid under
certain conditions. It has been argued both on general
principles\cite{thooftsmatrix} and in string theory\cite{nonlocal} 
that there are no truly local observables in quantum
gravity and that for this
reason the decomposition of the Hilbert space into sectors
inside and outside the black hole is invalid from the 
beginning. While this may indeed be true at some
fundamental level, the relevant question here is whether
local quantum field theory holds to a sufficiently good
approximation for points 3 and 4 to be valid. Since the
black hole  can be macroscopic and the curvature can be
very small compared with the string length or Planck
length, it is hard for me to see why the local field
theory approximation should fail in this regard. 
To postulate such a mysterious failure, when simpler scenarios
exist, seems to me uncalled for radicalism, although it is
a hypothesis favored by many physicists today.

\section*{Black hole entropy and surface states}

The previous arguments point to the conclusion that
black hole entropy is a measure of only those states
that can influence the outside of the black 
hole.\footnote{The case for a surface interpretation of black hole 
entropy has been made by various authors. In particular,
an article by Banks\cite{Banks:1994ph} (written before the age of D-branes)
makes the case with many of the same arguments as used here,
and the argument that the universality of black hole entropy  
(in spite of the {\it non}-universal history of the black hole)
arises from the universality of the near-horizon geometry
was made in a paper by Parentani and Piran\cite{P2}.}
These states must be associated with the presence of the
horizon, otherwise they would simply be counted as 
ordinary states of the exterior itself.  

One interpretation
of this ``surface entropy" is that it measures the information
in the entanglement of the vacuum across the horizon
(``entanglement entropy")\cite{sorkvac}. For
fields on a fixed background this is equivalent\cite{kabastra} 
to the
entropy of the thermal state (``thermal atmosphere") that
results when the state is restricted to the outside\cite{thooftwall}.
This entropy diverges, but gives something of the correct
order of magnitude if a Planck scale cutoff is imposed.

It is insufficient to consider fields on a fixed background
however. For one thing, although the contributions of quantum
fields can be thought of as ``loop corrections" to the black
hole entropy, there is also a classical contribution coming from 
the gravitational action itself. On can imagine an 
induced gravity scenario\cite{sakharov,tajind,froind1,froind2}, 
in which the entire gravitational action
is induced by matter, however there is still another problem:
for non-minimally coupled scalar fields or gauge fields, the 
entanglement entropy is not equal to the corresponding contribution
to the entropy computed from the induced gravitational action.  
It seems that the difference between these two entropies can be 
understood as a consequence of the fact that the background itself
varies when the temperature is varied\cite{frofurzel,frofur}.
Physically, this means that to understand the entropy
one must count states in the coupled matter-gravity vacuum. 

The large and universal number of states per unit
of surface area seems to be explained by the infinite redshift 
at the horizon: many states
at short distances near the horizon have the same, low, energy.
In fact, the number would appear to be {\it infinite} from 
perturbative counting, but the final count requires knowledge
of only the low energy effective gravitational action and the 
associated low energy Newton constant, as long as the spacetime 
curvature is small compared with Planck 
curvature\cite{susugl,tajind,larwil}. Although we 
are unable to compute the renormalized Newton constant
from quantum gravity, its (finite) value can be measured and
used in the entropy formula. 

\subsection*{Entanglement entropy and the generalized second law}
Sorkin proposed a derivation
of the generalized second law based on the entanglement 
interpretation of black hole entropy\cite{sorksm,sorksl}.
His idea was that the total entropy 
$S_{\rm outside}=S_{\rm horizon} +S_{\rm rest}$
of the reduced density
matrix outside the horizon receives a large universal contribution
$S_{\rm horizon}$ from the vicinity of the horizon and the rest
$S_{\rm rest}$ is  primarily just the ordinary entropy of a mixed 
state outside. Invoking the dynamical autonomy of the evolution
outside the horizon, Sorkin argues that $S_{\rm outside}$ cannot 
decrease, which amounts to the usual generalized second law 
provided $S_{\rm horizon}$ can be identified with the black hole
entropy.  This explanation of the generalized second law seems
so natural that it is hard to believe there is not some truth in it.
Unfortunately, as mentioned above, 
the entanglement interpretation of black hole
entropy does not seem to work, but perhaps this conclusion
is premature. 
Perhaps the black hole entropy could yet be 
understood in terms of entanglement entropy if,
as proposed in \cite{barfrozel}, the 
division of the system into inside and outside 
is referred to an intrinsic feature of the fluctuating
geometry such as the minimal throat area on some preferred
spacelike slice.

\section*{Objections}

Objections can be raised to the assertion that black 
holes have many more states than are counted by the 
black hole entropy. I believe that all of these objections 
are wrong, but it is challenging and instructive to try
to point to exactly where they are wrong. 
I will try to do so here with regard to several objections,
all but the first coming from string theory. 

\subsection*{Black hole pair creation amplitudes}

Semiclassical calculations of black hole pair creation
rates display a factor $\exp S_{\rm BH}$ which admits
the natural interpretation as a density of states 
factor\cite{paircreation}. This seems to lend solid 
support to the interpretation of $\exp S_{\rm BH}$ as
the number of states of the black hole. If the black
hole had more states, would they not contribute to
the pair creation rate?  This question has been discussed 
in the past, with conflicting
conclusions\cite{Banks:1993mi,Banks:1993is,Banks:1994ph,Gid:1994vj,Gid:1995qt}, 
and it deserves to be discussed further.
Here I will only state the reason for my belief that the
answer is no\footnote{I was
asked this question during my talk and had no quick answer.
After the talk Renaud Parentani suggested the following 
answer.}: Pair creation
is an exponentially suppressed tunneling process,
and any ``unnecessary" decoration of the black holes
would, it seems, be even more suppressed. 
All the extra internal states
are unnecessary decoration, and are 
therefore essentially irrelevant to the pair creation rates.  
 
\subsection*{String theory}

Calculations of black hole entropy in string theory
and its descendents have
been carried out in several contexts yielding agreement
with the Bekenstein-Hawking entropy. In all cases it 
appears that one is indeed counting all of the states
of the object identified with a black hole. Can this   be
compatible with the claim that black hole entropy  does
{\it not} count all of the states?   I will attempt to
argue that it can,  pointing to where one might find  the
other states. My attempts
are only partly successful, and are particularly weak
in the context of the AdS/CFT duality.

\subsubsection*{D-branes}

The entropy of certain near-extremal configurations of
D-branes has been found to agree with the semiclassical
entropy of the black hole configurations with the same 
set of charges 
(see for example\cite{DbranesMalda,DbranesPeet}). 
In the extremal case, for 
the supersymmetric BPS states, this is understood as a
consequence of the fact that the D-brane configuration
evolves into the black hole as the string coupling is 
increased from weak to strong, all the while maintaining
the supersymmetry. The enumeration of BPS states is 
independent of the coupling, hence the agreement in the 
count of states. In the D-brane picture there is nothing
corresponding to the inside of the black hole where extra
states can reside so, given the agreement with the black
hole entropy, how could a black hole have any more states?
For the BPS states the answer is simple:
the black hole also does not have any interior in the sense
that on a spacelike slice orthogonal to the timelike Killing
field the horizon is infinitely far away and has no
other side. 

The D-brane and black hole entropies also agree for 
near-extremal states however. In these cases, one can not
give such a simple answer. Imagine for instance  a
configuration that has been maintained at fixed energy above
extremality for a long time with  the help of an influx of
energy equal in magnitude to the Hawking flux. In the black
hole picture there is an arbitrarily large amount of
information stored in  the correlations between the inside
and outside of the black hole, so there must be a
correspondingly large number of states for the interior. In
the D-brane description however there is nothing that
corresponds to the interior. How could there be such a
drastic mismatch between the  total number of states in the
two descriptions and still be such agreement on not only the
entropy but also  the rate of Hawking emission (i.e. the
``greybody factors")?

I can give no really satisfying answer to this question.
Surely one has less control over the correspondence between
strong and weak coupling away from the BPS sector. It is
conceivable that the initial rates for Hawking radiation
agree but the details about the correlations that develop
over time do not match. In this scenario, there would simply
be more non-BPS states at strong coupling than there are at
weak coupling. This is not so hard to imagine, since in
going from weak to strong coupling the causal structure of
the background spacetime is distorted into that of a black
hole. An analogy that may be useful is the coupling constant
dependence of the state space  of electrons in an atom. At
sufficiently strong electric coupling the ground state
becomes unstable and electrons can be absorbed into the
nucleus, at which point the  nuclear Hilbert space comes
into play in resolving the physics. A strength of this
analogy is that  in the black hole case the ergoregion
inside the (non-extremal) horizon also manifests a kind of
instability of the ground state.  

\subsubsection*{AdS/CFT duality}

The near-horizon limit of the D-brane physics
led to the celebrated Maldacena conjecture, according to
which supergravity/string theory in an asymptotically
Anti-deSitter spacetime is equivalent to a superconformal
field theory on the conformal boundary of that 
spacetime\cite{magoo}. 
An example of this is the duality between superstring theory on
$AdS_5\times S^5$ and a $U(N)$ super-Yang-Mills
theory on $S^3\times R$, where $N$ is related to the 
string coupling $g_s$ the string length $\ell_s$
and the AdS radius $R$ by 
$R^4=4\pi g_s N \ell_s^4$, and the Yang-Mills
and string couplings are related by 
$g_{\rm YM}^2= 4\pi g_s$. 
There is much remarkable evidence in favor of 
the AdS/CFT duality, and no evidence against it to date.
Hence, for the sake of argument, let us suppose it is valid
and ask about the consequences for black holes.

In the $AdS_5\times S^5$ example it has been shown that the 
entropy of a
black hole which is large compared to the  AdS radius (and
is hence stable) is 3/4 of the  entropy of a thermal state
in the Yang-Mills theory  at weak 't Hooft coupling 
($g_{\rm YM}^2 N\ll 1$) at the corresponding Hawking temperature.
Moreover, there is reason to believe that the  entropy would
only change by a factor of order unity if the calculation
could be done at strong 't Hooft coupling (which is what is
required by for the case of large AdS radius).

We thus have a puzzle similar to that in the case of the
D-brane state counting, but now far from extremality. In the
Yang-Mills theory it seems there can be no missing states
corresponding to the degrees of freedom inside of the black
hole. The entropy of the thermal state simply counts all
states so, if the Maldacena conjecture  is really true, one
infers that there can be no independent degrees of freedom
inside the black hole. Can this conclusion be
evaded? 

\begin{figure}[hb]
\centerline{\epsfig{figure=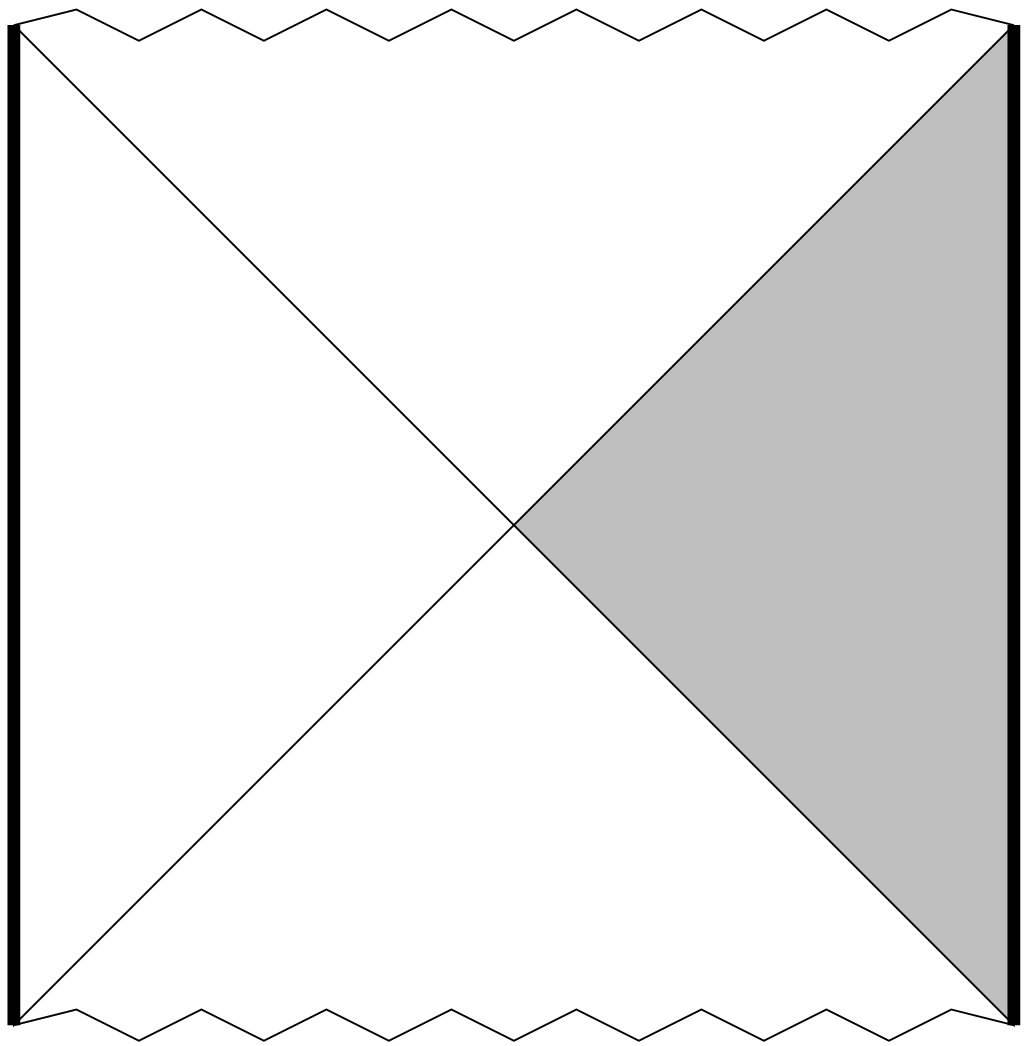,height=3cm}\hspace{4cm}
\epsfig{figure=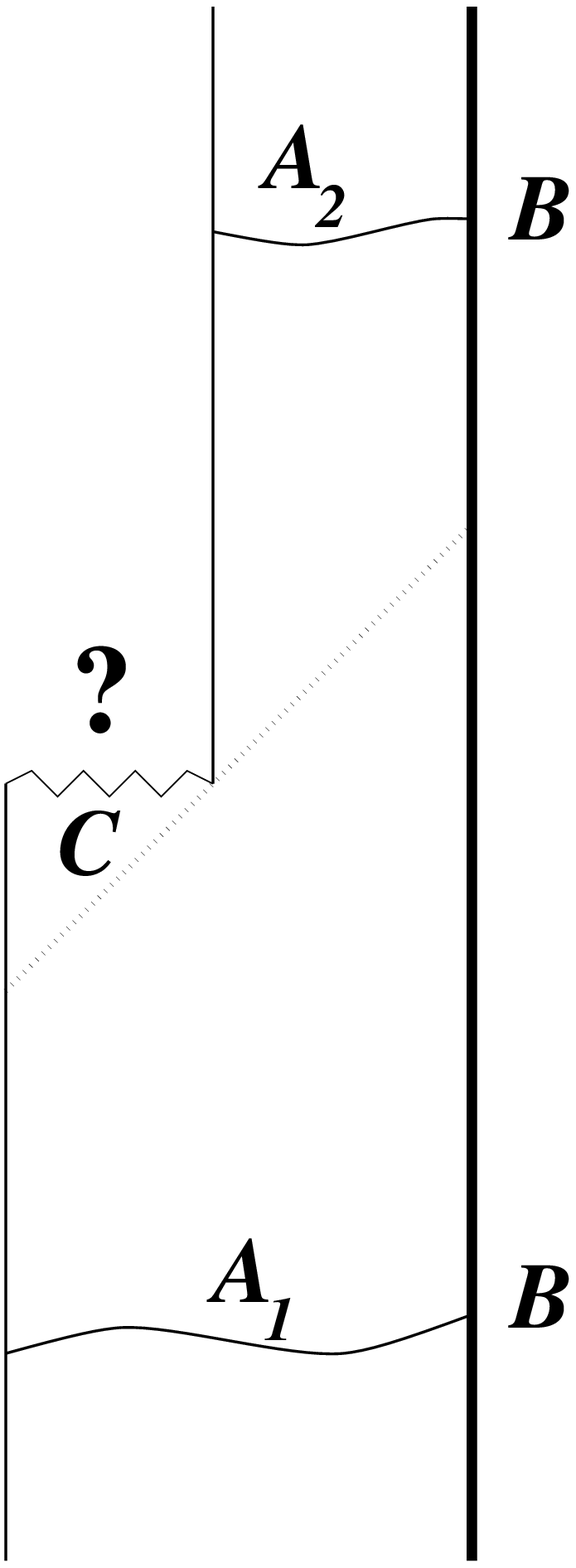,height=6cm}}
\vskip 3mm
\caption{({\bf a}) Schwarzschild-AdS spacetime. The conformal 
boundary is the pair of thick vertical lines. A single copy 
of the Yang-Mills
theory on the boundary may just correspond to the gravity
theory in the shaded wedge. ({\bf b}) Black hole formation 
and evaporation in Anti-de Sitter spacetime. The algebra 
of observables ${\cal A}_2$ may be a proper subalgebra of 
${\cal A}_2$.}
\label{adsfigs}
\end{figure}

A simple evasion is to suppose
that the equivalence conjectured by Maldacena actually
relates the supergravity observables only {\it outside} the 
horizon to the Yang-Mills observables in the boundary theory
(see Fig. \ref{adsfigs}(a)). 
This would be 
consistent with causality and would certainly explain why
all the states inside the black hole are not seen in 
the Yang-Mills theory. In fact, something like this seems
almost necessary in view of the fact that the full
Schwarzschild-AdS spacetime has a boundary with two
disconnected pieces, the dark vertical lines in Fig.
\ref{adsfigs}(a). What would be the role for the states in the 
Yang-Mills theory on the left if the one on the right 
already covered all the states inside the black hole? 

A different evasion is required if we consider not an
eternal black hole but rather a black hole,
small compared with the AdS radius,
that forms from collapse and then evaporates
(see Fig. \ref{adsfigs}(b)).
In this case the AdS/CFT duality presumably states that {\it
all}  observables in the spacetime have correspondents in
the CFT on the single boundary  component.  In particular, 
observables in the algebra $C$ localized behind the horizon 
(which is of course only defined relative to a particular 
state $|\psi\rangle$ of the CFT which corresponds to matter
collapsing to form a black hole) must be  contained within
the full algebra $B$ of observables in the CFT.  This in
itself is not mysterious, since the field equations allow us
to express any observable in $C$ as an observable in the
algebra $A_1$ localized at a spacelike slice before the
black hole ever formed.

The question of whether there are independent states of the
black hole interior is perhaps most sharply formulated here
as the question whether the black hole evaporation is
unitary from the viewpoint of the exterior\cite{cftunitarity}. 
Since the CFT
itself is unitary, the question amounts to whether 
the algebra  $A_2$ of observables on a spacelike 
slice after the black hole has evaporated completely 
is equal to $A_1$ or is rather a proper subalgebra of $A_1$.
In the latter case one would need also the observables in
the algebra $C$ behind the horizon to fill out the 
complete algebra. Moreover, causality would suggest that
$C$ and $A_2$ would commute relative to the state 
$|\psi\rangle$, that is, expectation values of the 
ideal generated by the commutator algebra $[C,A_2]$
would vanish. 

Many practitioners of duality have argued that the equality
$A_2=A_1$ is assured because the situations before collapse 
and after evaporation are similar: Anti-de Sitter spacetime
with some matter. In particular,  
the ``initial" configuration could have been the result
of a prior black hole formation and evaporation, or of
many cycles of formation and evaporation. If each black
hole has internal observables not captured on the outside
after the black hole is gone, then one seems to be requiring
that the CFT contains within it (relative for an appropriate
state) commuting subalgebras of observables   
corresponding to an infinite number of such black hole interiors, 
all of which commute with an algebra of outside observables
such as $A_2$. This 
requirement seems difficult to reconcile 
with reasonable expectations
about the number of states in the CFT at a given energy.

Can one really arrange a sequence of black hole
formations and evaporations where each black hole is made
from the Hawking radiation into which the previous black
hole evaporated? 
If not, then a state which produces many black
holes must contain to begin with energy corresponding 
to each black hole. In this case there there are perhaps 
more states so let us suppose, 
to be difficult, that one can indeed repeatedly refocus 
the Hawking radiation to form an endless cycle
of black holes  with a finite amount of
energy.  
In this case either the collection of 
commuting interior subalgebras exists, or one must
deny the independence of the interior observables.
Most string theorists support the second alternative.
I prefer the first since it requires only
nonintuitive behavior of the unfamiliar strongly coupled,
(astronomically) large N gauge theory, rather than 
gross violations of locality where they would not 
otherwise be expected.

\subsubsection*{Matrix Theory}

Matrix theory (a candidate for a nonperturbative
formulation of string theory) 
can purportedly describe formation and
evaporation of black holes, and the theory is manifestly
unitary.  There 
seems to be no room in matrix theory for
any states corresponding to the interior of a black hole,
left over after all particles in the Hawking radiation
have dissipated\cite{lenny}. I have not yet learned enough
about matrix theory to think carefully
about whether or not there is any
loophole through which this conclusion can be evaded.

\section*{Nature of the Bekenstein Bound}

The ``Bekenstein bound"\cite{bekbnd} 
on the entropy that can be associated with 
a closed 2-surface $\Sigma$ is
\begin{equation}
S_{\Sigma}\le \frac{A_{\Sigma}}{4\hbar G}.
\label{bekbound}
\end{equation}
This is (presently) a heuristic notion motivated by the 
generalized second law of thermodynamics as follows.
Suppose that by tossing in a suitable arrangement 
of matter the surface $\Sigma$ could be made to coincide with 
a slice of the horizon of a black hole. 
Then the entropy of that black hole would
be $A_{\Sigma}/{4\hbar G}$, which would violate the second
law unless the entropy $S_\Sigma$ associated with $\Sigma$ if
the extra matter is {\it not} tossed in is {\it less}, i.e. unless
the bound (\ref{bekbound}) holds. 

In describing the Bekenstein bound I was careful to
refer to $S_\Sigma$ as the entropy {\it associated with} $\Sigma$,
rather than the entropy {\it contained within} $\Sigma$, since the
meaning of the bound (\ref{bekbound}) inferred by the
black hole formation argument depends on the interpretation 
of the black hole entropy. If the black hole entropy is the 
logarithm of the number of states of the the black hole
including the interior states, then we infer a 
``volume bound" on
the entropy contained within $\Sigma$. If however, as
argued above, 
the black hole entropy reflects only those
states that can influence the exterior, then we infer only a 
``surface bound" on the surface states of $\Sigma$. 

I do not consider the volume bound interpretation to be viable.
Not only can it not be inferred from the second law with the surface
interpretation of black hole entropy, but it seems contradicted by 
the example used in the first section of this paper:  
since the volume of the region interior to 
the surface could be arbitrarily large it could 
contain an arbitrarily large amount of entropy.
It also suffers from a species problem, that is, 
the entropy inside could be arbitrarily large if the
number of independent fields in nature is arbitrarily large
(but see \cite{bekdowe} for another point of view). 
(On the other hand, if the number of species is
sufficient for an order unity violation of the bound,
then a black hole would be unstable to explosive 
evaporation on a timescale of order the light crossing 
time, and so the original rationale for the bound would be 
lost\cite{taj2b}.)
 
As an important side remark, note that the 
black hole formation argument
suggesting the bound
(\ref{bekbound}) does not apply to every 
closed 2-surface, since not every such surface 
can be made to coincide with a slice of the horizon 
of a black hole. Consider for instance an outer trapped
surface inside a black hole. The future pointing null
congruences orthogonal to this 
surface are converging on both sides,
whereas the horizon generators are always non-converging
according to the area theorem. For another example, consider
the intersection of the past light cones of two spacelike related
points $p$ and $q$. 
The future pointing null
congruences orthogonal to this intersection 
surface are converging (to $p$ and $q$)
on both sides. (This surface is not compact,
but one can build a compact 2-surface out of pieces like this.) 
The restriction on surfaces is certainly necessary
for the volume interpretation of the bound (although as 
discussed above I do not consider this interpretation to
be viable in any case), since
otherwise it is easy to find surfaces with arbitrarily little
area enclosing a large volume. For example, a trapped surface
near the singularity of a Schwarzschild black hole can have
arbitrarily small area and still bound a finite 
volume. For another example, 
one can make a spacelike 
surface of arbitrarily small area enclose any volume by
wiggling the surface ``up and down" in the timelike direction.

An interpretation of the bound (\ref{bekbound})
that is neither a volume nor
a surface interpretation has been proposed by Bousso\cite{raf}.
In this interpretation, $S_\Sigma$ is the entropy crossing
any segment of a null hypersurface, meeting
$\Sigma$ orthogonally, that is expanding towards $\Sigma$. 
The validity of this bound 
in a variety of contexts has been argued for in Ref. \cite{raf}. 

The volume bound interpretation of (\ref{bekbound}) suggests
the ``holographic principle"\cite{thooftbnd,holoprin} according 
to which
all the physics in the volume should be describable by a theory
on the bounding surface $\Sigma$. The surface bound interpretation
on the other hand does not have any holographic connotation.
Bousso suggests that his bound motivates a holographic 
priciple which refers to the null surface segments, but these
segments do not in general span the volume. 
It thus seems to me that the 
holographic principle, while it may be a property of quantum
gravity and/or of the AdS/CFT duality, is {\it not} logically
suggested by the Bekenstein bound.

\section*{Black hole entropy without black holes}

I have argued above
that black hole entropy is not 
determined by the number of internal states of the 
black hole, but rather by the number of states, associated
with the presence of the horizon, that can influence the 
outside world. This suggests that the notion of black
hole entropy should apply not just to black holes
but to any causal horizon. 

In fact, some approaches to computing the entropy
associated with horizons do yield the result $1/4$ per Planck
area of a Rindler horizon or a deSitter horizon, both
of which are observer dependent horizons. For example,
in a recent paper Carlip \cite{carlipcft} finds this
result from the representation theory of a conformal
subgroup of the diffeomorpism group associated with 
any (non-degenerate) Killing horizon, and he points out that
the Euclidean path integral approach also yields an entropy
for deSitter horizons\cite{gibbhawk}. Also,
the black hole pair creation probability is weighted by
$\exp(\Delta A_{\rm accel}/4)$ where $\Delta A_{\rm accel}$
is the associated increase of the area of an acceleration
horizon\cite{hawkhoroross}. This strongly 
suggests a state-counting role for the entropy of acceleration 
horizons, an idea which is further supported by calculations
relating transition amplitudes for particle creation
processes to the associated change of horizon 
area\cite{masspare}. (Ref. \cite{hawkhoro} argues that one 
should {\it not} attribute an entropy to the 
acceleration horizon because of its observer-dependent nature.
For the reason articulated in 
the concluding remarks, I do not subscribe to this viewpoint.)

As a more direct way to establish the validity
of horizon entropy without black 
holes, I will will now argue that there are 
general laws of horizon thermodynamics, strictly analogous 
to those for black holes,
for a class of causal horizons which
I will call ``partial event horizons".  
Recall that the global event horizon of
an asymptotically flat spacetime is the boundary of the past of 
future null infinity ${\cal I}^+$. I define similarly a
partial event horizon (PEH)
as the boundary $\partial I^-[p]$ of the past of 
a single point\footnote{One could of course
consider the boundary of the past of any subset
of ${\cal I}^+$.} 
$p\in{\cal I}^+$. In flat spacetime 
a PEH is just a Rindler (acceleration) 
horizon, and in an asymptotically flat spacetime
a PEH asymptotically approaches a Rindler horizon.

Although a PEH has cross sections with infinite area,  
it satisfies Hawking's classical area
theorem in the local sense that the 
expansion of its null 
generators is nowhere negative. The proof is similar to but
slightly simpler than that 
for the event horizon since the assumption of cosmic censorship
can be applied directly to rule out the possibility that
a null generator leaves the PEH before reaching ${\cal I}^+$.
Thus {\it changes} in the area are nonnegative, so a PEH
satisfies a classical ``second law of horizon mechanics".

A quasistationary region of a 
PEH also satisfies a ``first law of horizon 
mechanics" that is strictly analogous to the first law 
of black hole
mechanics $dM=(\kappa/8\pi)dA$.
This law for black holes can be understood in a quasi-local 
fashion, called the ``physical
process version" in Ref.\cite{waldredbook},
which applies to variations away
from a quasi-stationary configuration with approximate
horizon generating Killing field $\chi^a$.
In this setting $dM$ is interpreted  
as the flux $\int T_{ab}\chi^a d\Sigma^b$ of ``boost energy"
across the horizon or a part thereof.  A generic
PEH will possess many quasistationary regions,
to which the physical process version of first law will apply 
for the same reason as for black hole horizons. (The normalization
ambiguity of the boost Killing field scales both $dM$ and $\kappa$
in the same way, so the first law is independent of this 
ambiguity\cite{tos}.

Finally, as for the generalized second law, 
note that Sorkin's proposal for the 
origin of the generalized second law described above 
applies to any
causal horizon, and in particular it applies to a PEH.
Moreover, it seems that all gedanken experiments supporting
the generalized second law for quasistationary processes
involving black hole horizons would apply as well to 
quasistationary regions of PEH's.

\section*{Concluding Remarks}

What distinguishes a black hole horizon from a more general
causal horizon is that it is universally defined with reference
only to the global causal structure of the spacetime.
The absence of reference to particular observers or classes
of observers is thus its key distinguishing feature. In practice,
however, this universality is irrelevant. For example,
the universe may be spatially compact, and yet we have no
reservations in applying the laws of black hole thermodynamics
to approximately isolated ``black holes". It is always 
we who divide the system into the ``outside" and the 
``inside". It thus seems entirely natural that the notion of 
black hole entropy extends to general causal horizons.
This generalized notion of horizon entropy preserves the
the formula $S=A/4\hbar G$, whose universality is understood
as arising from the ultraviolet dominance of the 
``density of surface states", much as the universal form
of the short distance limit of quantum field correlations
is understood.

\section*{Acknowledgements}
I am grateful to numerous colleagues  
for even more numerous discussions on the topics discussed
here. This work was supported in part by the 
National Science Foundation under grants No. PHY98-00967 
at the University of Maryland and PHY94-07194 
at the Institute for Theoretical Physics.


\begin{references}

\bibitem{sorksm}
R.D.~Sorkin,
``The statistical mechanics of black hole thermodynamics,''
in {\it Black Holes and Relativistic Stars}, Chicago: The
University of Chicago Press, 1998, ch. 9, pp. 177-194;
gr-qc/9705006.

\bibitem{sorwaljiu}R.D.~Sorkin, R.M.~Wald, and Z.Z.~Jiu,  
``Entropy of self-gravitating radiation",
{\it Gen. Rel. Grav.} {\bf 13}, 1127 (1981).

\bibitem{thooftsmatrix}G.~'t Hooft,
``The Scattering matrix approach for the quantum black 
hole: An Overview,''
{\it Int. J. Mod. Phys.} {\bf A11}, 4623 (1996);
gr-qc/9607022.

\bibitem{nonlocal}D.A.~Lowe, J.~Polchinski, L.~Susskind, 
L.~Thorlacius and J.~Uglum,
``Black hole complementarity versus locality,''
{\it Phys. Rev.} {\bf D52}, 6997 (1995);
hep-th/9506138.

\bibitem{Banks:1994ph}
T.~Banks,
``Lectures on black holes and information loss,''
{\it Nucl.\ Phys.\ Proc.\ Suppl.\ } {\bf 41}, 21 (1995);
hep-th/9412131. 

\bibitem{P2}
R.~Parentani and T.~Piran,
``The Internal geometry of an evaporating black hole,''
{\it Phys.\ Rev.\ Lett.\ } {\bf 73}, 2805 (1994);
hep-th/9405007.

\bibitem{sorkvac}R.D.~Sorkin,
``On the Entropy of the vacuum outside a horizon,''
in B. Bertotti, F. de Felice, and A. Pascolini (eds.), 
{\it Tenth International Conference on General Relativity
and Gravitation, Contributed Papaers}, Vol. 2 
(Consiglio Nazionale Delle Ricerche, 1983), 734;
L.~Bombelli, R.K.~Koul, J.~Lee and R.D.~Sorkin,
``A Quantum source of entropy for black holes,''
{\it Phys. Rev.} {\bf D34}, 373 (1986).

\bibitem{kabastra}See, for example,
D.~Kabat and M.J.~Strassler,
``A Comment on entropy and area,''
{\it Phys. Lett.} {\bf B329}, 46 (1994);
hep-th/9401125.

\bibitem{thooftwall}G.~'t Hooft,
``On the Quantum structure of a black hole,''
{\it Nucl. Phys.} {\bf B256}, 727 (1985).

\bibitem{sakharov}A.D.~Sakharov,
``Vacuum quantum fluctuations in curved 
space and the theory of gravitation,''
{\it Sov. Phys. Dokl.} {\bf 12}, 1040 (1968). 

\bibitem{tajind}T.~Jacobson,
``Black hole entropy and induced gravity,''
gr-qc/9404039.

\bibitem{froind1}V.P.~Frolov, D.V.~Fursaev and A.I.~Zelnikov,
``Statistical origin of black hole entropy in induced gravity,''
{\it Nucl. Phys.} {\bf B486}, 339 (1997);
hep-th/9607104.

\bibitem{froind2}V.P.~Frolov and D.V.~Fursaev,
``Mechanism of generation of black hole entropy 
in Sakharov's induced gravity,''
{\it Phys. Rev.} {\bf D56}, 2212 (1997);
hep-th/9703178.

\bibitem{frofurzel} 
V.P.~Frolov, D.V.~Fursaev and A.I.~Zelnikov,
``Black hole entropy: Off-shell vs on-shell,''
{\it Phys. Rev.} {\bf D54}, 2711 (1996);
hep-th/9512184.

\bibitem{frofur}V.P.~Frolov and D.V.~Fursaev,
``Thermal fields, entropy, and black holes,''
{\it Class. Quant. Grav.} {\bf 15}, 2041 (1998);
hep-th/9802010.

\bibitem{susugl}L.~Susskind and J.~Uglum,
``Black hole entropy in canonical quantum 
gravity and superstring theory,''
Phys. Rev. {\bf D50}, 2700 (1994);
hep-th/9401070.

\bibitem{larwil} 
F.~Larsen and F.~Wilczek,
``Renormalization of black hole entropy and 
of the gravitational coupling constant,''
{\it Nucl. Phys.} {\bf B458}, 249 (1996);
hep-th/9506066.

\bibitem{sorksl}
R.D.~Sorkin,
``Toward an explanation of entropy increase in the 
presence of quantum black holes,''
{\it Phys. Rev. Lett.} {\bf 56}, 1885 (1986).

\bibitem{barfrozel}
A.O.~Barvinsky, V.P.~Frolov and A.I.~Zelnikov,
``Wave function of a black hole and the dynamical origin of 
entropy,''
{\it Phys. Rev.} {\bf D51}, 1741 (1995);
gr-qc/9404036.

\bibitem{Banks:1993mi}
T.~Banks and M.~O'Loughlin,
``Classical and quantum production of 
cornucopions at energies below $10^{18}$ GeV,''
{\it Phys.\ Rev.\ } {\bf D47}, 540 (1993);
hep-th/9206055.
 
\bibitem{Banks:1993is}
T.~Banks, M.~O'Loughlin and A.~Strominger,
``Black hole remnants and the information puzzle,''
{\it Phys.\ Rev.\ }{\bf D47}, 4476 (1993);
hep-th/9211030.

\bibitem{Gid:1994vj}
S.B.~Giddings,
``Comments on information loss and remnants,''
{\it Phys.\ Rev.\ } {\bf D49}, 4078 (1994);
hep-th/9310101.

\bibitem{Gid:1995qt}
S.B.~Giddings,
``Why aren't black holes infinitely produced?,''
{\it Phys.\ Rev.\ } {\bf D51}, 6860 (1995);
hep-th/9412159.

\bibitem{paircreation}
D.~Garfinkle, S.B.~Giddings and A.~Strominger,
``Entropy in black hole pair production,''
{\it Phys. Rev.} {\bf D49}, 958 (1994);
gr-qc/9306023.

\bibitem{DbranesMalda}J.M.~Maldacena,
``Black holes in string theory,''
hep-th/9607235.

\bibitem{DbranesPeet}A.W.~Peet,
``The Bekenstein formula and string theory (N-brane theory),''
{\it Class. Quant. Grav.} {\bf 15}, 3291 (1998);
hep-th/9712253.

\bibitem{magoo}O.~Aharony, S.S.~Gubser, J.~Maldacena, 
H.~Ooguri and Y.~Oz,
``Large N field theories, string theory and gravity,''
hep-th/9905111.

\bibitem{cftunitarity} 
See for example D.A.~Lowe and L.~Thorlacius,
``AdS/CFT and the information paradox,''
hep-th/9903237. 

\bibitem{lenny}L.~Susskind, private communication.

\bibitem{bekbnd}In the form given here the first reference
I know of is Ref. \cite{thooftbnd}. The name
``Bekenstein bound" is actually somewhat of a misnomer,
since the bound usually discussed by Bekenstein\cite{sebnd}
($S<2\pi ER$)
is actually a bound on the entropy given an ``energy" $E$ 
in a region of ``size" $R$. See however \cite{bekdowe}
for a discussion of the bound (\ref{bekbound}).

\bibitem{thooftbnd}G.~'t Hooft,
``Dimensional reduction in quantum gravity,'' in
{\it  Salamfestschrift},  edited by A.
   Ali, J. Ellis, S. Randjbar-Daemi, Singapore: 
   World Scientific, 1994;  
gr-qc/9310026. 

\bibitem{sebnd}J.D.~Bekenstein,
``A Universal upper bound on the 
entropy to energy ratio for bounded systems,''
{\it Phys. Rev.} {\bf D23}, 287 (1981).

\bibitem{bekdowe}J.D.~Bekenstein,
``Do we understand black hole entropy?,'' 
{\it The Seventh Marcel Grossmann Meeting on Recent Developments in
   Theoretical and Experimental General Relativity, Gravitation, and
   Relativistic Field Theories: Proceedings},
edited by  
R.T.~Jantzen, G. Mac Keiser, and Remo Ruffini (eds.), 
  Singapore: World Scientific, 1996; gr-qc/9409015.

\bibitem{taj2b}T.~Jacobson, to be published.

\bibitem{raf}R.~Bousso,
``A Covariant entropy conjecture,''
{\it JHEP} {\bf 07}, 004 (1999);
hep-th/9905177;
``Holography in general space-times,''
{\it JHEP} {\bf 06}, 028 (1999);
hep-th/9906022.

\bibitem{holoprin}L.~Susskind,
``The World as a hologram,''
{\it J. Math. Phys.} {\bf 36}, 6377 (1995);
hep-th/9409089.

\bibitem{carlipcft}
Carlip, S.,
``Entropy from conformal field theory at Killing horizons,''
gr-qc/9906126.

\bibitem{gibbhawk}
G.W.~Gibbons and S.W.~Hawking,
``Cosmological event horizons, thermodynamics, 
and particle creation,''
{\it Phys. Rev.} {\bf D15}, 2738 (1977).

\bibitem{hawkhoroross}
S.W.~Hawking, G.T.~Horowitz and S.F.~Ross,
``Entropy, area, and black hole pairs,''
{\it Phys. Rev.} {\bf D51}, 4302 (1995);
gr-qc/9409013.

\bibitem{masspare}
S.~Massar and R.~Parentani,
``Gravitational instanton for black hole radiation,''
{\it Phys. Rev. Lett.} {\bf 78}, 3810 (1997);
gr-qc/9701015;  ``How the change in horizon area drives 
black hole evaporation", gr-qc/9903027. 

\bibitem{hawkhoro}
S.W.~Hawking and G.T.~Horowitz,
``The Gravitational Hamiltonian, action, entropy and 
surface terms,''
{\it Class. Quant. Grav.} {\bf 13}, 1487 (1996);
gr-qc/9501014.

\bibitem{waldredbook}Wald, R.M.,
{\it Quantum Field Theory in Curved Spacetime and Black Hole
Thermodynamics}, Chicago: The
University of Chicago Press, 1994.

\bibitem{tos}T.~Jacobson,
``Thermodynamics of space-time: the Einstein equation of state,''
{\it Phys. Rev. Lett.} {\bf 75}, 1260 (1995);
gr-qc/9504004.

\end{references}
\end{document}